\documentclass[draft]{cimento}
\title{Mass formula for 2 dimensional flavorful mesons}
\author{Osamu Abe\from{ins:huea}
\thanks{e-mail: abeosamu@asa.hokkyodai.ac.jp}
 and Nobuaki Watanabe\from{ins:huea}
}
\instlist{\inst{ins:huea}
   Laboratory of Physics\\ 
   Asahikawa Campus, Hokkaido University of Education\\
   9 Hokumoncho, Asahikawa 070-8621, Japan
}
\PACSes{\PACSit{11.10.Ef}{Lagrangian and Hamiltonian approach}
        \PACSit{11.10.St}{Bound and unstable states; Bethe-Salpeter equations}
        \PACSit{11.15.Tk}{Other nonperturbative techniques}
}
\begin{document}
\maketitle
\hfill{}Preprint No.: HUEAP-15, hep-ph/0307229

\vspace{\baselineskip}
\begin{abstract}
We analytically and numerically investigate the  
't~Hooft equations,
the lowest order mesonic Light-Front Tamm-Dancoff equations for 
$\rm SU(N_C)$ and $\rm U(N_C)$gauge theories, generalized to flavor 
non singlet mesons.
We find the wave function can be well approximated by new basis 
functions and obtain an analytic and an empirical formulae for the 
mass of the lightest bound state. Its value is consistent with the 
precedent results.
\end{abstract}

The light front Tamm-Dancoff (LFTD) method 
\cite{lftd,brodsky98,burkardt} has been 
introduced as an alternative tool to lattice gauge theory to 
investigate relativistic bound states nonperturbatively.
In the LF coordinate, the physical vacuum is equivalent to the 
bare vacuum, since all constituents must have 
non-negative longitudinal momenta defined by
$k^+=(k^0+k^3)/\sqrt{2}$.  Because of this simple structure of 
the true vacuum, we can avoid the serious problems which appeared in 
the Tamm-Dancoff (TD) approximation \cite{tammdancoff} in the equal 
time frame. Therefore, the TD approximation is 
commonly used in the context of the LF quantization.

The techniques have been developed 
\cite{ma,mo,harada94,sugihara,abe97,abe99} 
for solving LFTD equations in several models such as the massive 
Schwinger model \cite{coleman}, which is the extension of the simplest 
(1+1)-dimensional QED${}_2$ \cite{schwinger}. Bergknoff 
\cite{bergknoff} first applied LFTD approximation to the massive 
Schwinger model. 
In the most of above references, as they concentrated mainly on taking 
account of the higher Fock state contributions systematically in the 
context of LFTD approximation, they analyzed LFTD equations assuming
that all the masses of quarks are degenerated in order to avoid 
complexities of numerical treatment. 
Mo and Perry \cite{mo} and Harada and coworkers \cite{harada94} 
introduced basis functions to treat the massive Schwinger model in the
context of LFTD approximation. One of the present author \cite{abe99} 
generalized their basis functions. But, all the basis functions are
applicable only in the case where all quark masses are degenerated.

In the real world, as six quark have their inherent masses, there are 
many mesons consist of quark and anti-quark with different masses.
In this short note, we will attempt to generalize  basis functions 
so as to treat the masses of mesons consist of different flavors 
with different masses. 
We will neglect the contributions from higher Fock states, then 
we are led to the generalized 't~Hooft-Bergknoff-Eller equation 
\cite{bergknoff,thooft}
\begin{eqnarray}
   \left[M^2 -\frac{m_f^2-1}{x}-\frac{m_{f'}^2-1}{1-x}\right]
   \Phi(x)
   =-\wp\!\! \int_{0}^{1}\frac{\Phi(y)}{(y-x)^2}dy
    +\alpha\int_{0}^1\Phi(y)dy.
\label{generalthooft}
\end{eqnarray}
Here, parameter $\alpha$ specifies the model under consideration, 
{\it i.e.}, $\alpha =0$ for $\rm SU({\mathit N_C})$ and 
$\alpha =1$ for $\rm U({\mathit N_C})$, $\wp$ stands for the 
Hadamard's finite part. 
$M$ is the dimensionless meson mass, $m_f$ and $m_{f'}$ are 
the dimensionless quark mass of flavor f and f'. They are related to 
the coupling constant $g$ and bare masses $\bar{M}$ and $\bar{m}_f$ as follows:
\begin{equation}
   M^2=\frac{2\pi N_C\bar{M}^2}{(N_C^2+\alpha-1)g^2},\;
   m_f^2=\frac{2\pi N_C\bar{m_f}^2}{(N_C^2+\alpha-1)g^2}.
\end{equation}

One of present authors (O.A.) \cite{abe99} pointed out that the wave 
function can be expanded in terms of $(x(1-x))^{\beta_n+j}$ in case 
$m_{f'}=m_f$. Here, $\beta_n$ is the $(n+1)$-th smallest positive 
solution of Eq. (\ref{beta}) which will be given bellow.
Main interest in the present paper is to extend 
the basis functions so that we can treat mesons with 
$m_f \neq m_{f'}$. One may expect it is enough to extend above basis 
function to $x^{\beta_n+j}(1-x)^{\beta'_n+j}$. As we will see shortly,
it is not the case.

At first, according to 't~Hooft \cite{thooft}, we put
\begin{equation}
   \Phi(x)=x^\beta(1-x)^{\beta '}.
\label{thooftansatz}
\end{equation}
The most singular part of the left hand side of 
Eq. (\ref{generalthooft}) at the end point $x=\epsilon$  is given by 
$-(m_f^2-1)\epsilon^{\beta-1}$.
One of the right hand side of Eq. (\ref{generalthooft}) is given by
\begin{equation}
  -\beta\pi\cot(\pi\beta) \epsilon^{\beta -1}.
\end{equation}
Here, we have used
\begin{eqnarray}
   \wp\int_{0}^{1} \frac{y^a(1-y)^b}{(y-x)^2}
   &=&B(a-1,b+1) F(2,1-a-b;2-a;x)\nonumber\\
   & &-\pi\cot(\pi a)\left\{ ax^{a-1}(1-x)^b
        - b x^a (1-x)^{b-1} \right\}\nonumber\\
   &\equiv& f_{ab}(x),
\end{eqnarray}
where $B$ is the beta function and $F$ denotes the Gauss's 
hypergeometric function. Thus, we are led to
\begin{equation}
   m_f^2-1+\beta\pi\cot(\pi\beta)=0.
\label{beta}
\end{equation}

Analogously, we also have at another end point $x=1-\epsilon'$,
\begin{equation}
   {{m_f}'}^2-1-\beta' \pi\cot(\pi\beta' )=0.
\label{betaprime}
\end{equation}
Here, we have used $f_{ab}(x)=f_{ba}(1-x)$.
If we assume $\beta \simeq O(m_f)$ for small $m_f$, we obtain
\begin{eqnarray}
   \beta&=&\frac{\sqrt{3}}{\pi}m_f+{\cal O}(m_f^2),\\
   \beta'&=&\frac{\sqrt{3}}{\pi}{m_f}'+{\cal O}({{m_f}'}^2).\nonumber
\end{eqnarray}
We multiply both sides of Eq. (\ref{generalthooft}) by $\Phi(x)$ and
integrate them, we have
\begin{eqnarray}
   M^2 B(1+2\beta,1+2\beta')&=&(m_f^2-1) B(2\beta,1+2\beta')
      +(m_{f'}^2-1)B(1+2\beta,2\beta)\\
    & &-I(\beta,\beta',\beta,\beta')
      + \alpha B(1+2\beta,1+2\beta').\nonumber
\label{massequation1}
\end{eqnarray}
Here, 
\begin{eqnarray}
 I(a,b,c,d)&\equiv& 
     \int_{x=0}^{1}\wp \!\!\int_{y=0}^{1}
   \frac{y^{a}(1-y)^{b}x^{c}(1-x)^{d}}
   {(y-x)^2}dy  dx\\
   &=& -\pi a\cot (\pi a ) B(a+c,b+d)\nonumber\\
   & &+ \pi(a+b )\cot (\pi a )
           B(1+a+c,b+d)\nonumber\\
   & &+ B(-1+a,1+b) B(1+c,1+d) \times\nonumber\\
   & &  {}_3F_2( 2,1 - a - b ,1 + c ;
       2 - a ,2 + c  + d ;1).\nonumber
\label{ap8.3}
\end{eqnarray}
In the above equation, ${}_3F_2$ denotes the generalized hypergeometric
function. Thus, for small $m_f$ and small $m_{f'}$, we are led to
\begin{eqnarray}
   M^2=\alpha+\frac{\pi}{\sqrt{3}}(m_{f}+m_{f'})+{\cal O}(m_f^2,
          m_fm_{f'},{m_{f'}}^2).
\label{lowestordermass}
\end{eqnarray}

As a next step, we consider a higher order correction to Eq. 
(\ref{thooftansatz}). The most general form of the wave function
is given by
\begin{equation}
   \Phi(x)=\lim_{N\to\infty}\sum_{n_1=0}^N\sum_{j_1=0}^{N-n}
           \sum_{n_2=0}^N\sum_{j_2=0}^{N-n}
          C_{n_1}^{j_1}{}_{n_2}^{j_2}
          x^{\beta_{n_1}+j_1}
          (1-x)^{\beta'_{n_2}+j_2},
\label{generalwave}
\end{equation}
where $\beta_n$ and $\beta_n'$ are $(n+1)$-th smallest positive 
solution of Eq. (\ref{beta}) and Eq. (\ref{betaprime}), 
respectively. The reason why we cannot introduce a term other than 
one in Eq. (\ref{generalwave}) will be presented later.

If we substitute Eq. (\ref{generalwave}) into Eq. 
(\ref{generalthooft}), we have, at the end point $x=\epsilon$,
\begin{eqnarray}
 0&=&-\sum_{n_1=0}^{\infty}(m_f^2-1+\pi\beta_{n_1}\cot\pi\beta_{n_1})
     \sum_{n_2,j_2}
          C_{n_1}^{0}{}_{n_2}^{j_2}\epsilon^{\beta_{n_1}-1}\\
  & &+\sum_{n_1,J=0}^{\infty}\left[
   M^2\sum_{k=0}^J\sum_{n_2,j_2=0}^{\infty}
          C_{n_1}^{J-k}{}_{n_2}^{j_2}\frac{(-\beta'_{n_2}-j_2)_k}{k!}
       \right.\nonumber\\
   & &-(m_f^2-1)\sum_{k=0}^{J+1}\sum_{n_2,j_2=0}^{\infty}
          C_{n_1}^{J+1-k}{}_{n_2}^{j_2}\frac{(-\beta'_{n_2}-j_2)_k}{k!}
      \nonumber\\
   & &-(m_{f'}^2-1)\sum_{k=0}^J\sum_{n_2,j_2=0}^{\infty}
          C_{n_1}^{J-k}{}_{n_2}^{j_2}\frac{(-\beta'_{n_2}-j_2+1)_k}{k!}
       \nonumber\\
   & &-\sum_{k=0}^{J+1}\sum_{n_2,j_2=0}^{\infty}\pi(\beta_{n_1}+J+1-k)
          \cot\pi\beta_{n_1}
      C_{n_1}^{J+1-k}{}_{n_2}^{j_2}\frac{(-\beta'_{n_2}-j_2)_k}{k!}
      \nonumber\\
   & & +\left. \sum_{k=0}^J\sum_{n_2,j_2}\pi(\beta'_{n_2}+j_2)
          \cot\pi\beta_{n_1}
      C_{n_1}^{J-k}{}_{n_2}^{j_2}\frac{(-\beta'_{n_2}-j_2+1)_k}{k!}
      \right]\epsilon^{\beta_{n_1}+J}\nonumber\\
   & &+\sum_{k=0}^{\infty}
       \sum_{n_1,j_1,n_2,j_2}C_{n_1}^{j_1}{}_{n_2}^{j_2}
      \Biggl[
      B(\beta_{n_1}+j_1-1, 1+\beta'_{n_2}+j_2)\times\nonumber\\
   & &   \frac{(2)_k(1-\beta_{n_1}-j_1-\beta'_{n_2}-j_2)_k}
           {(2-\beta_{n_1}-j_1)_k k!}
      \nonumber\\
   & & -\alpha \delta_{k0}
      B(1+\beta_{n_1}+j_1,1+\beta'_{n_2}+j_2)
      \Biggr]   \epsilon^k.\nonumber
\label{endpointeq}
\end{eqnarray}
Here, $\displaystyle (a)_n\equiv {\Gamma (a+n)\over \Gamma (a)}$ 
is the Pochhammer symbol. The first line in Eq. (\ref{endpointeq}) 
vanishes automatically, because of the definition of $\beta_n$.

Now, it becomes clear that Eq. (\ref{generalwave}) is the most general 
form. 
Suppose that we introduce the term like $c x^\gamma(1-x)^{\gamma'}$ 
with $\gamma \neq \beta_n+j$, then it requires 
\begin{equation}
  0 = c (m_f^2-1+\pi\gamma\cot\pi\gamma)\epsilon^{\gamma -1}
\end{equation}
to hold. Thus, the coefficient $c$ should vanish. Analogously, 
if $\gamma' \neq \beta'_n+j$ then $c=0$.

We also have similar equation to Eq. (\ref{endpointeq}) at another 
end point $x=1-\epsilon'$. If we truncate Eq. (\ref{generalwave}) to 
given finite $N$. The total number of free parameters is $(N+1)^2(N+2)^2/4$.
On the other hand, if we require Eq.(\ref{endpointeq}) and similar equation 
to hold upto of order $O(\epsilon^{\beta_{N-1}})$ or 
$O(\epsilon'{}^{\beta'_{N-1}})$, 
we have $N(N+3)$ independent equations. Thus, we cannot solve
the equations in general. We have to reduce the degree of freedom.
We put 
\begin{equation}
     C_{n_1}^{j_1}{}_{n_2}^{j_2}=
        \delta_{n_10}\delta_{n_20}\delta_{j_1j_2}d_{j_1}
       +\delta_{j_10}\delta_{j_20}e_{n_1n_2},
\end{equation}
that is we assume 
\begin{equation}
   \Phi(x)=
    \sum_{j=0}^N d_j x^{\beta_0+j}(1-x)^{\beta'_0+j}
    +\sum_{n1,n_2=0}^N e_{n_1n_2}
    x^{\beta_{n_1}}(1-x)^{\beta'_{n_2}},
\label{newansatz}
\end{equation}
with $d_{00}=1$ and $e_{00}=0$.

For a given $m_f$ and $m_{f'}$, we put $M^2=M_i^2$.
We can then solve Eq.(\ref{endpointeq}) for ${d_j}$ and 
$e_{n_1n_2}$ in terms of $M_i$.
We thus obtain the $M_i$ dependent truncated wave function, say, 
$\Phi(x;M_i)$. We can calculate a new mass eigenvalue $M_{i+1}$ using 
this wave function as 
\begin{eqnarray}
M_{i+1}^2={<\Phi(M_i)|H|\Phi(M_i)>\over <\Phi(M_i)|\Phi(M_i)>}.
\end{eqnarray}
We can use Eq.(\ref{lowestordermass}) as $M_0^2$. 
For $N\le 3$, mass eigenvalue $M^2$ converges in 5 iterations. 
For $N=3$ and $0<m_f,\; m_{f'}\le 0.1$, we 
obtain $M^2$'s by the use of {\it Mathematica}.
The results are summarized in Tables \ref{table1} and 
\ref{table2}. We can fit them by polynomials:
\begin{eqnarray}
M^2(\alpha=0,m)&=&1.8139(m_f + m_{f'}) + 0.892(m_f + m_{f'})^2 
   + 0.008m_f \, m_{f'} \\
   & &+  0.041(m_f + m_{f'})^3 
   - 0.18m_f \, m_{f'}(m_f + m_{f'}) + \cdots\nonumber
\end{eqnarray}
\begin{eqnarray}
M^2(\alpha=1,m)&=&1 + 1.8092(m_f + m_{f'}) + 0.497(m_f + m_{f'})^2 
   + 1.564m_f \, m_{f'} \\
   & &+ 0.95(m_f + m_{f'})^3 
   - 4.40m_f \, m_{f'}(m_f + m_{f'}) + \cdots.\nonumber
\label{ourresult}
\end{eqnarray}

We cannot proceed this procedure beyond $N=3$, because we cannot
calculate ${}_3F_2(2,a,b;c,d;1)$ in desired precision.
The results with $m_{f'}=m_f$ are consistent with 
the previous results in \cite{abe99}.

Finally, we will discuss a so-called ``2\% discrepancy'' problem.
In ${\rm U}(N_C)$, single flavor model, we expect the dimensionless 
meson mass squared can be expanded as follows, 
\begin{equation}
   M^2=1+b_1m_f +b_2m_f^2+\cdots.
\end{equation}
By the use of the bosonization method, Banks and his coworker 
\cite{banks} found $b_1=2\exp (\gamma_E)=3.56214\cdots$. Bergknoff 
\cite{bergknoff}, however, found $b_1=2\pi/\sqrt{3}=3.62759\cdots$.
Two results differ each other by 2\%. Our result, 
Eq. (\ref{ourresult}), is consistent with Bergknoff's result rather 
than Banks {\it et al.}'s. We may expect that $b_1=2\pi/\sqrt{3}$ is 
the minimum value in the context of 't~Hooft-Bergknoff-Eller equation 
and higher Fock sector should be included to solve the problem.

\section*{Acknowledgements}
This work was supported by the Grants-in-Aid for Scientific Research of 
Ministry of Education, Science and Culture of Japan (No. 14540001).

\begin{table}
\begin{center}
\caption{Numerical results for bound state mass $M^2$ in 
$\rm SU(N_C)$ model as a function of quark masses $m_f$ and 
$m_{f'}$.}
\label{table1}
\begin{tabular}{rr|rrrrrr}
\hline
\multicolumn{2}{c}{} &\multicolumn{6}{c}{$m_{f'}$}\\
\cline{3-8}
  \multicolumn{2}{c}{        } &    0.01 &    0.02 &    0.04 &
     0.06 &    0.08 &   0.10\\
\hline
\hline
      &0.01 &0.03663 &0.05522 &0.09293 &0.13136 &0.17051 &0.21036\\
      &0.02 &0.05522 &0.07398 &0.11205 &0.15084 &0.19034 &0.23056\\
      &0.04 &0.09293 &0.11205 &0.15083 &0.19033 &0.23055 &0.27149\\
$m_f$ &0.06 &0.13136 &0.15084 &0.19033 &0.23056 &0.27147 &0.31312\\
      &0.08 &0.17051 &0.19034 &0.23055 &0.27147 &0.31317 &0.35546\\
      &0.10 &0.21036 &0.23056 &0.27149 &0.31312 &0.35546 &0.39866\\
\hline
\end{tabular}
\end{center}
\end{table}
\begin{table}
\begin{center}
\caption{Numerical results for bound state mass $M^2$ in 
$\rm U(N_C)$ model as a function of quark masses $m_f$ and 
$m_{f'}$.}
\label{table2}
\begin{tabular}{rr|rrrrrr}
\hline
\multicolumn{2}{c}{} &\multicolumn{6}{c}{$m_{f'}$}\\
\cline{3-8}
   \multicolumn{2}{c}{}
            &0.01    & 0.02   & 0.04   & 0.06   & 0.08   & 0.10\\
\hline
\hline
      &0.01 &1.03660 &1.05506 &1.09233 &1.13012 &1.16847 &1.20738\\
      &0.02 &1.05506 &1.07387 &1.11159 &1.14988 &1.18867 &1.22799\\
      &0.04 &1.09233 &1.11159 &1.15040 &1.18954 &1.22922 &1.26940\\
$m_f$ &0.06 &1.13012 &1.14988 &1.18954 &1.22965 &1.27018 &1.31122\\
      &0.08 &1.16847 &1.18867 &1.22922 &1.27018 &1.31154 &1.35348\\
      &0.10 &1.20738 &1.22799 &1.26940 &1.31122 &1.35348 &1.39622\\
\hline
\end{tabular}
\end{center}
\end{table}
\end{document}